\documentstyle [a4,12pt,psfig] {article}

\begin{document}
\newcommand{\nwc}{\newcommand}
\nwc{\hyp} {\hyphenation}
\nwc{\be}  {\begin{equation}}
\nwc{\ee}  {\end{equation}}
\nwc{\ba}  {\begin{array}}
\nwc{\ea}  {\end{array}}
\nwc{\bdm} {\begin{displaymath}}
\nwc{\edm} {\end{displaymath}}
\nwc{\bea} {\be\ba{lcl}}
\nwc{\eea} {\ea\ee}
\nwc{\bda} {\bdm\ba{lcl}}
\nwc{\eda} {\ea\edm}
\nwc{\bc}  {\begin{center}}
\nwc{\ec}  {\end{center}}
\nwc{\ds}  {\displaystyle}
\nwc{\bmat}{\left(\ba}
\nwc{\emat}{\ea\right)}
\nwc{\non} {\nonumber}
\nwc{\hph} {\hphantom}
\nwc{\qq}  {\qquad}
\nwc{\lra} {\longrightarrow}
\nwc{\ra}  {\rightarrow}
\nwc{\Ra}  {\Rightarrow}

\nwc{\rng} {\rangle}
\nwc{\lng} {\langle}

\nwc{\lmt} {\longmapsto}
\nwc{\sq}  {\sqrt}
\nwc{\prl} {\partial}
\nwc{\prlb}{\bar{\prl}}
\nwc{\fc}  {\frac}
\nwc{\kr}  {\kern}
\nwc{\iy}  {\infty}
\nwc{\ol}  {\overline}
\nwc{\hm}  {\hspace{3mm}}
\nwc{\Er}   {{\rm E}}
\nwc{\tr} {\rm Tr}
\nwc{\lf} {\left}
\nwc{\ri} {\right}
\nwc{\lm} {\limits}
\nwc{\lb} {\lbrack}
\nwc{\rb} {\rbrack}
\nwc{\ov} {\over}
\nwc{\ovx}{\over\textstyle}
\nwc{\til}{\tilde}
\nwc{\pr} {\prime}
\nwc{\st} {\strut}
\nwc{\vs}  {\vskip}
\nwc{\noa} {\noalign}
\nwc{\scr}  {\scriptstyle}
\nwc{\tx}  {\textstyle}
\nwc{\scs} {\scriptscriptstyle}
\nwc{\Th} {\Theta}
\nwc{\th} {\theta}
\nwc{\vth} {\vartheta}
\nwc{\eps}{\epsilon}
\nwc{\si} {\sigma}
\nwc{\Gm} {\Gamma}
\nwc{\gm} {\gamma}
\nwc{\bt} {\beta}
\nwc{\La} {\Lambda}
\nwc{\la} {\lambda}
\nwc{\om} {\omega}
\nwc{\Om} {\Omega}
\nwc{\dt} {\delta}
\nwc{\Si} {\Sigma}
\nwc{\Dt} {\Delta}
\nwc{\al} {\alpha}
\nwc{\vph}{\varphi}
\nwc{\Sis} {\Si^{(s)}}
\nwc{\sis} {\si^{(s)}}
\nwc{\Sims}{\Sigma^{(-s)}}
\nwc{\Us}  {U^{(s)}}
\nwc{\sipl} {\sigma^+}
\nwc{\simi} {\sigma^-}
\nwc{\sd}  {\dot s}
\nwc{\Id}  {{\bf 1}}
\nwc{\Sc}  {{\cal S}}
\nwc{\Cr}  {{\cal R}}
\nwc{\Cd}  {{\cal D}}
\nwc{\Oc}  {{\cal O}}
\nwc{\Cc}  {{\cal C}}
\nwc{\Of}  {{\cal O}_f}
\nwc{\Oft} {{\cal O}_{f_2}}
\nwc{\Ofo} {{\cal O}_{f_1}}
\nwc{\Pc}  {{\cal P}}
\nwc{\Mc}  {{\cal M}}
\nwc{\Ec}  {{\cal E}}
\nwc{\Fc}  {{\cal F}}
\nwc{\Hc}  {{\cal H}}
\nwc{\cK}  {{\cal K}}
\nwc{\Fcp} {{\cal F}^\pr}
\nwc{\Hcp} {{\cal H}^\pr}
\nwc{\Xc}  {{\cal X}}
\nwc{\Gc}  {{\cal G}}
\nwc{\Zc}  {{\cal Z}}
\nwc{\Nc}  {{\cal N}}
\nwc{\fca} {{\cal f}}
\nwc{\xc}  {{\cal x}}
\nwc{\Ac}  {{\cal A}}
\nwc{\Bc}  {{\cal B}}
\nwc{\Uc} {{\cal U}}
\nwc{\Vc} {{\cal V}}
\nwc{\rh}  {{\hat r}}

\nwc{\nnn} {\nonumber \vspace{.2cm} \\ }
\nwc{\hb}  {\bar h}
\nwc{\xb}  {\bar x}
\nwc{\yb}  {\bar y}
\nwc{\zb}  {\bar z}
\nwc{\wb}  {\bar w}
\nwc{\Ob}  {\bar O}
\nwc{\Yb}  {\bar Y}
\nwc{\cv} {{\vec c}}
\nwc{\ph} {{\hat p}}
\nwc{\wh} {{\hat w}}
\nwc{\qh} {{\hat q}}
\nwc{\xh} {{\hat x}}
\nwc{\php} {{\hat p}^\pr}
\nwc{\whp} {{\hat w}^\pr}
\nwc{\qhp} {{\hat q}^\pr}
\nwc{\xhp} {{\hat x}^\pr}
\nwc{\bp}  {b^\pr}
\nwc{\ft} {{\til f}}
\nwc{\ct} {{\til c}}
\nwc{\tit}{{\til t}}
\nwc{\wt} {{\til w}}
\nwc{\Pb} {{\bf  P}}
\nwc{\Zb} {{\bf  0}}
\nwc{\Eb} {{\bf  1}}
\nwc{\Qb} {{\bf  Q}}
\nwc{\Bb} {{\bf  B}}
\nwc{\Tb} {{\bf  T}}
\nwc{\qqd}  {\qq .}
\nwc{\diag} {{\rm diag}}
\nwc{\inv}  {{\rm inv}}
\nwc{\mod}  {{\rm mod}}
\nwc{\clr}  {{\rm cl}}
\nwc{\qur}  {{\rm qu}}
\nwc{\Pro}   {{\rm P}}
\nwc{\Yr}   {{\rm Y}}
\nwc{\sof}  {{\st 1 \ov \sqrt{|\Of|}} }
\nwc{\sofo} {{\tx 1 \ovx \sqrt{\tx |\Ofo|}} }
\nwc{\soft} {{\tx 1 \ovx \sqrt{\tx |\Oft|}} }
\nwc{\hal} {\frac{1}{2}}
\nwc{\tpi}  {2\pi i}
\nwc{\gpr}  {g^\pr}
\nwc{\zot}  {z_{12}}
\nwc{\zbot} {\zb_{12}}
%-----------------------------------------------------------------------
%abbreviations for number sets:
%
\def\KK{{\rm I\kern -.2em  K}}
\def\NN{{\rm I\kern -.16em N}}
\def\RR{{\rm I\kern -.2em  R}}
\def\ZZ{{\small{\rm Z}\kern -.34em Z}}
\def\ZZZ{{\small{\rm Z}\kern -.5em Z}}
\def\QQ{{\rm \kern .25em
             \vrule height1.4ex depth-.12ex width.06em\kern-.31em Q}}
\def\CC{{\rm \kern .25em
             \vrule height1.4ex depth-.12ex width.06em\kern-.31em C}}

%commands added in Heidelberg
\newcommand{\Ga}{\Gamma}
\newcommand{\bul}{\bullet}
\newcommand{\bull}{$\bul$}
\newcommand{\Rr}{{\rm R}}
\newcommand{\Lr}{{\rm L}}

\renewcommand{\theequation}{\thesection.\arabic{equation}}

\newcommand{\req}[1]{(\ref{#1})}
\newcommand{\sect}[1]{ \section{#1} \setcounter{equation}{0} }
\newcommand{\ve}{\left( \begin{array}{r}}
\newcommand{\ev}{\end{array} \right)}
\newcommand{\arr}{\left( \begin{array}{rrrr}}
\newcommand{\arrr}{\left( \begin{array}{rrrrrr}}
\newcommand{\eqr}{ \begin{eqnarray}}
\newcommand{\rqe}{ \end{eqnarray}}
\newcommand{\eq}{\begin{equation}}
\newcommand{\qe}{\end{equation}}
\newcommand{\mreq}[1]{\req{#1:start} -- \req{#1:end}}

% Mathematical macros

\newcommand{\half}{\mbox{$\frac{1}{2}$}}

\newcommand{\xmu}{\mbox{$ X^{\mu} $}}
\newcommand{\wmu}{\mbox{$ w^{\mu} $}}
\newcommand{\pleft}{\mbox{$ p_{L} $}}
\newcommand{\pright}{\mbox{$ p_{R} $}}
\newcommand{\lpleft}{\mbox{$ \hat{p}_{L} $}}
\newcommand{\lpright}{\mbox{$ \hat{p}_{R} $}}
\newcommand{\emui}{\mbox{$ e^{\mu}_{\ \ i} $}}
\newcommand{\esmui}{\mbox{$ e_{\mu}^{\ast\ i} $}}
\newcommand{\lat}{\mbox{$ \Lambda $}}
\newcommand{\dlat}{\mbox{$ \Lambda^{\ast} $}}
\newcommand{\nlat}{\mbox{$ \Lambda_{\cal N} $}}
\newcommand{\mhat}{\mbox{$ \hat{m} $}}
\newcommand{\nhat}{\mbox{$ \hat{n} $}}
\newcommand{\es}{\mbox{$ e^{\ast} $}}
\newcommand{\et}{\mbox{$ e^{T} $}}
\newcommand{\qs}{\mbox{$ Q^{\ast} $}}
\newcommand{\qt}{\mbox{$ Q^T $}}
\newcommand{\qi}{\mbox{$ Q^{-1} $}}
\newcommand{\ws}{\mbox{$ W^{\ast} $}}
\newcommand{\wi}{\mbox{$ W^{-1} $}}
\newcommand{\oms}{\mbox{$ \Omega ^{\ast} $}}
\newcommand{\omt}{\mbox{$ \Omega ^T $}}
\newcommand{\omi}{\mbox{$ \Omega ^{-1} $}}
\newcommand{\cm}{\mbox{$ c_{\hat{m}}$}}
\newcommand{\cn}{\mbox{$ c_{\hat{n}}$}}
\newcommand{\sip}{\mbox{$ \sigma ^{(+)}_i $}}
\newcommand{\simm}{\mbox{$ \sigma ^{(-)}_i $}}
\newcommand{\sjm}{\mbox{$ \sigma ^{(-)}_j $}}
\newcommand{\skpm}{\mbox{$ \sigma ^{(\pm)}_k $}}
\newcommand{\slpm}{\mbox{$ \sigma ^{(\pm)}_l $}}
\newcommand{\fs}{\mbox{$ \hat{f^{\ast}}$}}
\newcommand{\ket}[1]{\mbox{$ \mid #1 >$}}
\newcommand{\h}{\mbox{$\chi$}}
\newcommand{\oddz}{$O(d, d;\ZZ)\ $}
\newcommand{\sldz}{$SL(d;\ZZ)\ $}
\newcommand{\sqt}{\sq^T}
\newcommand{\ig}{\mbox{$ g^{-1}$}}
\newcommand{\mdsg}{modular discrete symmetry group\ }
\newcommand{\tgroup}{\mbox{$ {\cal G}_{\cal N} $}}
\newcommand{\ogroup}{\mbox{$ {\cal G}_{\cal O} $}}
\newcommand{\gmo}{\mbox{$ \tilde{\cal G}_{\cal O} $}}
\newcommand{\ms}{\mbox{$ {{\cal M}_{\cal O}} $}}
\newcommand{\p}{\partial}
\newcommand{\f}{\hat{f}}
\newcommand{\w}{\hat{v}}
\newcommand{\m}{\hat{p}}
\newcommand{\igpb}{\frac{1}{g + b}}
\newcommand{\igmb}{\frac{1}{g - b}}
\newcommand{\su}{$\spadesuit$}

\begin{titlepage}

\title{Yukawa Couplings for Bosonic $Z_N$ Orbifolds: \\
      Their Moduli and Twisted Sector Dependence\thanks{Supported by
      Deutsche Forschungsgemeinschaft}}

\author{{\sc S. Stieberger$^1$,}
        \and {\sc D. Jungnickel$^{1,2}$,} \and
        {\sc J. Lauer$^3$} \and
        {\ \ \ \ \ and \sc \ M. Spali\'nski$^1$}\thanks{
        Alexander von Humboldt Fellow.
        On leave from the Institute of Theoretical Physics,
        Warsaw University.}\\ \\
       {\em $^1$Physik Department} \\
       {\em Technische Universit\"at M\"unchen} \\
       {\em D--8046 Garching, Germany}  \\ \\
       {\em $^2$Max-Planck-Institut f\"ur Physik} \\
       {\em ---Werner-Heisenberg-Institut---}\\
       {\em P.O. Box 401212, D--8000 M\"{u}nchen, Germany}\\ \\
       {\em $^3$Institut f\"ur Theoretische Physik} \\
       {\em Universit\"at Heidelberg} \\
       {\em Philosophenweg 16} \\
       {\em D--6900 Heidelberg, Germany}}

\date{}
\maketitle

\begin{picture}(5,2.5)(-300,-560)
\put(12,-100){MPI--Ph/92-24}
\put(12,-115){TUM--TH--134/91}
\put(12,-130){HD--THEP--92-15}
\put(12,-148){March 1992}
\end{picture}

\thispagestyle{empty}

\begin{abstract}

The three point correlation functions with twist fields are determined
for bosonic $Z_N$ orbifolds. Both the choice of the modular background
(compatible with the twist) and of the (higher) twisted sectors involved
are fully general.
We point out a necessary
restriction on
the set of instantons contributing to twist field correlation functions
not obtained in previous calculations.
Our results show that the theory
is target space duality invariant.

\end{abstract}
\end{titlepage}

\sect{Introduction}

Toroidal orbifolds
are a class of exactly soluble string compactifications that
provide the possibility of studying string models both from the
phenomenological and more formal points of view
\cite{DHVW}--\nocite{IMNQ,DFMS}\cite{HV}. This class
shares some interesting features with other types of string compactification.
As an example we mention
the concept of marginal deformations of a conformal field
theory \cite{DVV}. At the string theory level the deformation parameters
({\em moduli}) describe
the K\"ahler and complex structure of the target space for
some connected class of models.
Generically there appears a modulus whose real part is the square of
the scale factor of the compact space.
On the other hand they amount to flat directions in the scalar potential of the
low energy effective theory.

An object of prime interest in any phenomenological
study of string compactification are the Yukawa couplings. Their knowledge
is clearly
required if one wishes to explain the quark and lepton masses, mixing
angles and so forth from first principles.
In string theory these couplings receive moduli dependent corrections due
to the presence of worldsheet instantons \cite{DSWW,DFMS}.
This non--perturbative effect yields exponentially suppressed factors that are
governed by the instanton actions.
The latter depend linearly on the moduli. This property has attracted
interest already in the past, since it naturally leads to hierarchies of
Yukawa couplings \cite{DFMS,HV}. One of the
apparent shortcomings of the presently existing string theories
is the fact that they allow for a
{\em multitude} of string vacua
which are locally parametrized in terms of the moduli.

An analysis of the spectrum of simple string compactifications already
reveals
discrete symmetries in moduli space, which are more precisely referred to as
target space symmetries
\cite{DVV,SAS}, \cite{SW}--\nocite{GRV,DHS,LMN1,LMN2,MS1,MS2}\cite{EJN}.
A prominent example is provided by the {\em duality} map which
relates models
whose (finite) target space volumes are in inverse proportion.
To limit the range of values for the moduli to
a fundamental region representing inequivalent models, such discrete symmetries
must be applied. They may also hint at the existence of a fundamental
length in string theory \cite{GRV,LMN1,GV}. Furthermore, assuming that these
string target
space symmetries carry over to the low energy supergravity effective
Lagrangian, one may use them to restrict the form of the scalar
potential \cite{FLST}.
%the task of determining the scalar potential is enormously facilitated
%\cite{FLST}.

Hence it is important to know whether duality symmetries are also
respected by the interactions of compactified
heterotic string models. Many examples are known where this
can be answered in the affirmative \cite{DVV,LMN1,LMN2},
\cite{AO}--\nocite{EJLM,EJL}\cite{COPG}. At present it would be
rather surprising if there were interactions which break duality
invariance.
Although no rigorous proof exists to date,
one is tempted to consider for instance duality
invariance as a {\em consistency check} on string calculations.
Indeed, a major motivation of the investigation in \cite{S}
was a recent computation of
general twisted sector Yukawa couplings in \cite{BKM}. If correct,
one would  have to conclude that the duality symmetry
is broken by twisted sector interactions.
As shown below, this is not in fact the case.

Correlation functions of
twist fields from the lower twisted sectors
were first computed by Dixon,
Friedan, Martinec, and Shenker \cite{DFMS}, and (independently) by Hamidi
and Vafa \cite{HV}. In both cases the twisted Narain compactification
\cite{IMNQ,N,NSW} consisted of a purely metrical background.
Next, the antisymmetric background was accounted for:
the case of a two--dimensional target space
was treated in \cite{LMN2}, and subsequently
the annihilation couplings \cite{EJLM}
and particular Yukawa couplings \cite{EJL} were derived given a general
background in higher dimensions.

The extension to an arbitrary choice of higher sector twist fields
was initiated in \cite{BKM} for the case of a purely metrical
background. Six--dimensional $Z_N$ Coxeter orbifolds \cite{IMNQ}
 provide
numerous examples some of which have recently been worked out in  \cite{CGM}.
In our communication this issue is resumed.
All three point correlation functions for bosonic $Z_N$ orbifolds
are determined. Throughout the calculation we allow for an
arbitrary constant background provided that it commutes with the twist
operation. Some results related to higher sector twist field
configurations do not agree with the findings of \cite{BKM}.
These discrepancies show up if the
{\em complete} set of global monodromy conditions
is taken into account which characterize the admissible
worldsheet instantons.
Since the requisite calculations
are somewhat involved,
this note will concentrate on presenting our results
and pointing out some subtleties, which do not appear in the cases
considered in the basic work \cite{DFMS}
but lie at the origin of
the difference between our results and those of \cite{BKM}.
A more complete presentation is deferred to \cite{ejss}.

As suggested above, the various couplings will put
the duality and other background symmetries to the test.
In section 4 we touch on the representation of a duality map on
twisted sectors which is intimately related to the form of three point
correlation functions. We stress that
according to our results duality invariance is maintained
as regards the most general Yukawa interactions in the presence of a
metrical and an antisymmetric background tensor.

\section{The four point function}

The class of models considered here may be described by the action \cite{NSW}
\eq
S(G,B) = \frac{1}{2 \pi} \int  d \bar{z} \ \bar{\p} X^j (G+B)_{jk}\p X^k
       \ . \label{action}
\qe
The coordinate fields $X^j(z,\bar z)$
describe the embedding of the string in
the $d$--di\-men\-sio\-nal compact target space. From now on $d$ will be
treated
as an even number, the constant
target space metric $G$ and the antisymmetric background tensor $B$
refer to the coordinate basis. As discussed
\cite{GRV,EJLM} $G$ can be
arranged to be proportional to $\Eb$ via a coordinate
transformation. From now
 on $G={1\ov2}\Eb_d$ is taken.

The construction of $Z_N$ orbifolds, as described in
\cite{DHVW,IMNQ},
proceeds by imposing twisted boundary conditions on the fields $X$:
\eq
 X(e^{2\pi i}z,e^{-2\pi i}\bar z)=(\Theta X)(z,\bar z) + 2\pi w\ .
\label{bocon}
\qe
Here the translation vector $w$ belongs to a lattice
$\Lambda$ which describes the compact toroidal target space.
The {\em twist} $\Th$
 generates the $Z_N$ point group symmetry of
the orbifold. Its eigenvalues are the phases
\be
u_j = \exp (2\pi i \alpha_j) \ \ \ \ j = 1, \ldots , d/2\ ,
\label{evs}
\ee
and their complex conjugates such that $N \alpha_k \in \ZZ$\ .
The embedding $X$ is split into a classical {\em instanton} solution
and a quantum fluctuation part:
\be
X^j(z,\bar z)=X_{\clr}^j(z,\bar z)+X_\qur^j(z,\bar z)\ ,
\ \ \ j=1, \ldots,d\ ,
\ee
where $X_\clr^j$ solves the equation of motion
$\p \bar \p X^j(z,\bar z)=0$ for \req{action}.
In the neighbourhood
of a twist field located at the origin the classical
field must obey \req{bocon} (as is required for the full field)
which is why the quantum part only feels the {\em local} monodromy:
\be
X_\qur^j(e^{2\pi i}z,e^{-2\pi i}\bar z)=(\Theta X_\qur)^j(z,\bar z)\; .
\ee

In order for \req{action} to be consistent with \req{bocon}
the twist $\Th$ must be chosen ortho\-gonal w.r.t. the metric $G$.
Likewise consistency requires $B$ to obey (see \cite{EJLM})
\eq [B, \Th] = 0 \ . \qe

Our first task is to calculate the four twist field correlation function

\eq
Z_{\{f_{i}\}}(x,\bar x)=\lim_{|z_{\infty}| \rightarrow
\infty}|z_{\infty}|^{4h_l}\ \lng \sigma_{f_1}^{-k}(0,0)\sigma_{f_2}^{k}(x,\bar
x)\sigma_{f_3}^{-l}(1,1)\sigma_{f_4}^{l}(z_{\infty},\bar z_{\infty}) \rng
\label{4pf}
\qe
from which the various three point functions follow by factorization.
Here $\sigma_f^k$ denotes the twist field associated with the
fixed point $f$
in the sector twisted by $\Th^k$, i.e., $(1-\Th^k) f %\tilde{f}
\in \Lambda$ (according to (\ref{bocon}) the rescaled vector $2\pi f$
provides a fixed point of the toroidal target space  under $\Th^k$).
Stated more precisely, the elements of the {\em conjugacy class}
\eq
C(k;f) = [\Th^k,(1-\Th^k)(f+\Lambda)]    \label{conj}
\qe
determine the boundary conditions (\ref{bocon})
of nearby coordinate fields.
This twist field has the (anti--) conformal dimension
\eq
     h_k = \bar h_k = \sum_{j=1}^{d/2} h_{k_j}\ , \label{confd}
\qe
where
\eq h_{k_j} = \frac{1}{2} k_j(1-k_j)\ ,\ \ \ {\rm with}\quad
    k_j = \lb k \al_j\rb \in (0,1)\ .
\qe
For non--prime orbifolds one must of course construct
twist invariant linear combinations of the basic twist fields $\sigma_f^k$,
but then the calculation reduces again to computing (auxiliary)
correlation functions of the type \req{4pf}\,(see \cite{EJLM}).
Note furthermore that twisted sectors which contain invariant
subspaces (i.e. some $k_j$ vanish) are not dealt with in this article. This is
done in \cite{ejss}.

The method used here to calculate couplings is that of Dixon et al.
\cite{DFMS} and is based on the path integral formalism.
Since the action \req{action} is bilinear in the fields $X^j$,
the path integral representation of the twisted vacuum
configuration \req{4pf} splits into two factors, one
coming  from the {\em classical} instanton solutions $X_\clr(z,\bar z)$,
the other describing the {\em quantum} fluctuations around them.
This means
\be
Z_{\{f_{i}\}}(x,\bar x) = Z_\qur(x,\bar x)\ \sum_{X_\clr}
e^{-S [{1\ov2}\Eb_d,B;X_\clr]}\ ,
\label{z}
\ee
where the instanton solutions $X_\clr^j$ are subject to the full set
of boundary
conditions \req{bocon} imposed by the four twist fields in \req{4pf}.

In the sequel we prefer to have the rotational twist $\Th$
in its simplest $2\times2$ block diagonal form.
If we then pass to new complex coordinate fields
\eq
Y^j(z,\bar z) := X^{2j-1}(z,\bar z) + i X^{2j}(z,\bar z)\ ,
\qquad j = 1,\ldots,d/2\ ,
\qe
the twist operation becomes $Y^j\mapsto u_j Y^j$.

The instanton solutions $Y_\clr$ display branch
point singularities as a consequence of (\ref{bocon})
(e.g. $\p Y^j(e^{2\pi i} z) = u_j \p Y^j(z)$).
Hence their form is \cite{BKM}
\be
\ba{clccr}
\ds{\p Y^j_\clr(z)} &=& \ds{b^j \omega_{k_j,l_j}(z)\ ,} \\[2mm]

\ds{\bar \p Y^j_\clr(\bar{z})} &=&
\ds{c^j \bar{\omega}_{1-k_j,1-l_j}(\bar{z})\ ,}
\label{insta}
\ea \ee
where
\eq \ba{lcr}
 \ds{\omega_{k_j,l_j}(z)} &=&
 \ds{z^{-k_j}(z-x)^{k_j-1}(z-1)^{-l_j}(z-z_{\infty})^{l_j-1}\ ,} \nnn
 \ds{\omega_{1-k_j,1-l_j}(z)} &=& \ds{z^{k_j-1}(z-x)^{-k_j}
 (z-1)^{l_j-1}(z-z_{\infty})^{-l_j}\ ,} \ea \label{cutdifferentials}
\qe
$$ j = 1,\ldots,d/2\ .$$
The complex coefficients $b^j$ and $c^j$ normalize the instantons and may be
 recovered
by line integration from the holomorphic, (antiholomorphic)
functions $\p Y$, $\bar\p Y$ so that
the global monodromy conditions \cite{DFMS} will be satisfied:
\eq \begin{array}{rcl}
 \ds{\triangle_{C_a} Y^j_\clr}&\equiv&\ds{ \oint_{C_a} dz \,\p Y_\clr^j
 +\oint_{C_a} \, d\bar z \bar \p
 Y_\clr^j = 2\pi v_a^{j}} \nnn
 \ds{\triangle_{C_a} \bar Y^j_\clr} &\equiv&\ds{ \oint_{C_a} dz
 \, \p \bar Y^j_\clr +
 \oint_{C_a} d \bar z \, \bar \p \bar Y^j_\clr =2\pi \bar v_a^{j}}\ ,
 \end{array} \label{glomono}
\qe
$$j=1,\cdots,d/2\ .$$
Candidates for the contours $C_a$ appearing here are any
possible zero net twist loops
which exist in the case at hand.
The complex vectors $v_a$ belong to the lattice cosets which appear in the
space group product of the conjugacy classes
of the encircled twist fields (see (\ref{conj})).
To determine the normalization of the
instanton solutions one may consider, as in \cite{DFMS,BKM}, the pair of
independent loops $C_1$, $C_2$ in fig. $1$. While for $k=l$ this exhausts
the content
of the global monodromy conditions that matter for
the instanton sum, in general
these conditions are more restrictive.
We will therefore also have to include the loop $C_3$ of fig. $1$.

\ \\
\ \\
%\begin{figure}[h]
\psfig{figure=stem.ps,height=95mm,width=115mm}
%\end{figure}
\ \\
\ \\

\begin{minipage}[t]{5.4in}
{Fig. 1. The zero net twist loops $C_1$, $C_2$, $C_3$ are depicted
for
the choice $p=2$, $q=1$ of the winding numbers of $C_2$.
By virtue of ${\rm SL}(2,\CC)$--invariance the twist field locations
$z_1=0$, $z_2=x$, $z_3=1$, and $z_4= z_\infty$ are adopted (cf. (2.7)).}
\end{minipage}
\ \\
\ \\
\ \\
By multiplying the space group conjugacy classes \req{conj}
corresponding
to the fields encircled
by $C_a$  one finds that the shift $v_a$ must satisfy
\bea
\ds{v_1} &=& \ds{(1-\Th^k)(f_{21}+\lambda_1)\ ,} \nnn
\ds{v_2} &=& \ds{(1-\Theta^{pk})(f_{23}+\mu)\ ,} \\[2mm]
\ds{v_3} &=& \ds{(1-\Th^l)(f_{43}+\lambda_2)\ ,} \nonumber
\label{glomon}
\eea
where $\ds{\lambda_1,\lambda_2,\mu \in \Lambda}$ and $f_{ab}$
$\equiv$ $f_a - f_b$.
The winding numbers $p$, $q$
of $C_2$ about $z_2 =x$, $z_3=1$ are related via $pk=ql$ (see fig. 1).

The global monodromy condition for $C_3$ implies that
\be v_3= - v_1\ . \label{v1v3}
\ee
The union $C_1+C_3$ which encircles all
twist fields can of course be shrunk to $0$. % as is evident from fig. 1.
This forces the global monodromy shift
to vanish. We then arrive at the  {\em space group selection rule}
\eq
(1-\Th^k) f_{21}+(1-\Th^l)f_{43} = (1-\Th^k)T_1+(1-\Th^l)T_2\ ,
\label{srule}
\qe
with $T_1, T_2 \in \Lambda$.
It must be interpreted as follows: If there is
a pair $T_1$, $T_2$ $\in\Lambda$
satisfying this rule
then the correlation function \req{4pf} will not vanish.
In order to determine the available global shifts \req{glomon}
we combine \req{v1v3} and \req{srule} into
\be
(1-\Th^k)\lambda_1+(1-\Th^l)\lambda_2=-(1-\Th^k)T_1-(1-\Th^l)T_2\ .
\ee
Solving this condition for $\lambda_1$ yields
\be
\lambda_1 \in  \til \La \equiv -T_1 + {\ds{1-\Th^l }\ovx
 1-\Th^\phi}\La\ ,
\label{range}
\ee
$\phi$ being the greatest common divisor of $k$ and $l$. Of course
$\lambda_2\in\Lambda$ is then assured. Conversely, solving for
$\lambda_2$ we get
\be
\lambda_2 \in -T_2 + {\ds{1-\Th^k}\ovx 1-\Th^\phi} \La
\label{range2}
\ee
and again $\lambda_1\in \Lambda$ will ensue.

Note that the global monodromy sets \req{glomon}
are unambiguously determined,
in the sense that they do not depend on the choice of $T_1$, $T_2$,
as long as the latter solve \req{srule}.
Indeed, a different choice of $T_1$, $T_2$ (while keeping
$f_{21}$, $f_{43}$ fixed) can be readily absorbed by suitable shifts
of the $\Lambda$ factors present in \req{range}, \req{range2}.
Another ambiguity is met if the fixed point $f_{21}$
($f_{43}$) is displaced by a lattice vector ---
this will be compensated by shifting $T_1$ ($T_2$) by the same lattice
vector.

The classical action evaluated for the instanton solutions will be denoted by
$S[G, B; v_1,v_2]$,
since it depends on the vectors $v_1, v_2$ via the normalization
coefficients $b^i$ and $c^i$ which were determined by (\ref{glomono}) for
 $j=1,2$. This calculation is
somewhat lengthy, but straightforward.
Only the results of the factorization procedure will be presented
in section 3
(for further details see \cite{S}).

The quantum part of the correlation function may be calculated by the stress
tensor method \cite{DFMS,BKM}. In this way one arrives at
\eq
Z_{\{f_{i}\}}(x,\bar x) = \nu\ |x|^{-4h_k}
\lf[\prod_{j=1}^{d/2}\fc{(1-x)^{-l_j (1-k_j )}
(1-\xb )^{-k_j (1-l_j )}}{I_j (x,\xb )}\ri] \hspace{-1.0cm}
\ds{\sum_{\hspace{1.2cm} v_1 \in (1-\Th^k) (f_{21} + \tilde{\Lambda})
    \atop \hspace{1.2cm} v_2 \in
(1-\Th^{pk})(f_{23} + \Lambda)} \hspace{-0.8cm}
e^{-S [{1\ov2}\Eb_d,B;v_1,v_2]}} \; .
\label{4pfres}
\qe
where $\tilde{\Lambda}$ denotes the {\em restricted}
lattice from \req{range} and the normalization constant
\be
\nu =(2\pi)^{d\ov2} V_\Lambda
\label{norm}
\ee
guarantees a canonical normalization of the twist field two point functions
($V_\Lambda$ is the volume of a unit cell of $\Lambda$).

The function $I_j (x,\xb )$ is given by
\eq
I_j(x,\xb ):=  % \fc{\sin (pk_j\pi )}{\pi}
     J_{2j}\ol{G}_{1j}(\xb )H_{2j}(1-x)+
     J_{1j}G_{2j}(x)\ol{H}_{1j}(1-\xb ) \ ,
\qe
with (the hypergeometric function is abbreviated by $F(a,b;c;x)$)
\be
\ba{rcl}
 \ds{G_{1j}(x):= F(k_j,1- l_j;1;x)}&,&
 \ds{G_{2j}(x):= F(1-k_j,l_j;1;x)} \nnn
 \ds{H_{1j}(x):= F(k_j,1-l_j;1+k_j-l_j;x) }&,&
 \ds{H_{2j}(x):= F(1-k_j,l_j;1+l_j-k_j;x) } \nnn
 \ds{J_{1j}:=\frac{\Ga(k_j) \Ga(1-l_j)}{\Ga(1+k_j-l_j)}}&,&
 \ds{ J_{2j}:=\frac{\Ga(1-k_j)\Ga(l_j)}{\Ga(1+l_j-k_j)}}\ .
\ea
\ee

The fact that the restricted lattice $\tilde \La$ appears in \req{4pfres}
is very important -- were it not the case $Z_{\{f_j\}}$ would not be symmetric
 under permutations of the twist fields. The expression for the four
point correlation function given in \cite{BKM}
is in fact not symmetric
because the global monodromy about the loop $C_3$ was not taken into account
there. If however

\eq
\det \fc{1-\Th^l}{1-\Th^\phi} = 1\ ,
\label{det}
\qe
then this loop does not give any new information, and
\cite{BKM} contains the correct result.
This is obviously the case for prime orbifolds, as well as for general $Z_N$
orbifolds if $l$ divides an element of $k+N\ZZ$.
However even if (\ref{det}) is met,
the results given in \cite{BKM}
must be used carefully due to the lack of permutation symmetry
mentioned above.

%\newpage
\sect{The three point functions}

There are two kinds of three point functions to consider: the coupling of two
twisting states to an untwisted sector state,
and the coupling of three twisting states.

The coupling of two twisted sector states to an untwisted one is defined as
\eq
C^l_{f_4,f_3;p,w}:=\lim_{|z_{\iy}|\ra \iy} |z_{\iy}|^{4h_l}  \,
    \lng V_{-p,-w}^{\inv}(0,0)\si_{f_3}^{-l}(1,1)
    \si_{f_4}^{l}(z_{\iy},\zb_{\iy}) \rng \ . \label{ssv}
\qe
Here $V_{p, w}^{\inv}$ denotes a twist invariant vertex operator, and
$h_k$ is the conformal dimension (\ref{confd}) of the twist field in the
 sector twisted by
$\Th^k$.

The couplings \req{ssv} must appear in the $s$--channel factorization
of the four point function.
 From space group arguments and $SL(2,\CC )$--invariance we already learn
that
the operator product expansion of two twist fields takes the form
\eq
\sigma^{-k}_{f_1}(0,0)\sigma^{k}_{f_2}(x,\bar{x})
   =\frac{1}{N} \hspace{-.6cm}\sum_{p\in\La^{\ast}\atop
          \hspace{0.6cm} w \in(1-\Th^k)(f_{21}+\La)}
   \hspace{-.5cm}x^{h-2h_k}\
   \bar{x}^{\bar{h}-2h_k}\ C^k_{f_2,f_1;p,w}\
   V_{p,w}^{\inv}(x,\bar{x})+ \ldots\ ,   \label{ope}
\qe
where the powers $h, \bar h$ in the local expansion
denote the left and right conformal dimensions of the vertex
operator $V_{p, w}^{\inv}$:
\bdm
\ba{lclccrcclccr}
    & \ds{h} &=& \ds{\frac{1}{2}P_\Rr^t P_\Rr} &,&
    \ds{\bar h} &=& \ds{\frac{1}{2} P_\Lr^t P_\Lr\ } \\[5mm]
    {\rm with} & \ds{P_\Rr} &=& \ds{p-Bw +{\ds w\ov 2}}
    &,& \ds{P_\Lr} &=& \ds{p-Bw-{\ds w\ov 2}} \ .
\ea
\edm
Inserting (\ref{ope}) into the four point function (\ref{4pf}) one easily finds
for infinitesimal $x, \bar x$
\eq
Z_{\{f_{i}\}}(x,\bar x) = \frac{1}{N}\hspace{-.6cm}\sum_{p \in \La^{\ast}
  \atop \hspace{.6cm} w \in
  (1-\Th^k) (f_{21} + \tilde{\La})} \hspace{-.5cm}
  C^k_{f_2,f_1;p,w}\ C^l_{f_4,f_3;-p,-w}
\; |x|^{-4 h_k} x^h\  \bar x^{\bar h} + \ldots\ . \label{fac}
\qe

This has to be compared with the corresponding limit of the
correlation function \req{4pfres}. To this end one has to perform
a Poisson resummation in the winding $v_2$.
%he four point function given in
%ection 2 can be examined in the limit of small $x, \bar x$. The resulting
%xpression can then be compared with \req{fac}. This is easily done due to the
Due to the manifestly
factored form of the $k$ and $l$ dependences
in the coefficients of \req{fac} one readily gets
\be
   C_{f_2,f_1;p,w}^k = g_k(P_\Lr,P_\Rr) \
   e^{-2\pi i f_2^t p}\ \
   e^{i \pi p^t \frac{1}{1-\Theta^k}w}\
   \dt_{w\in (1-\Th^k)(f_{21}+\Lambda)}\ ,
   \label{svs}
\ee
where
\bdm
   g_k(P_\Lr,P_\Rr) = \sqrt{N}\ \lf[ \frac{1}{N}
   \ri]^{\frac{1}{2}(P_\Rr^2+P_\Lr^2)}
   \; \prod_{j=1}^{N-1}|1-\theta^j|^{\frac{1}{2}(P_\Rr^t
   \Th^{jk} P_\Rr+P_\Lr^t \Th^{jk} P_\Lr)}\ ,\ \ \ \theta=e^{\fc{2\pi i}{N}}\ .
\edm
A different derivation of this result has been achieved with the
help of the mode operator method in \cite{EJLM}.
We stress that the small
$|x|$ limit of $Z_{f_i}(x,\xb)$ arrived at in \cite{BKM} does not lend itself
to an $s$--channel factorization \cite{S}.

Another point is, that strictly speaking, one must perform a summation over
the {\em orbits}
${\cal O}_{f_j}=\{\Th^n f_j\ \mod \Lambda;\ 0\le n< k\}$
of the two fixed points entering \req{svs} to get the physical coupling of
the twist invariant ground states $\Sigma_{f_{2,1}}^{\pm k}$.

The Yukawa couplings are defined by
\eq Y_{f_a,f_b,f_c}^{k,l}\equiv \lim_{|x|\ra \iy}
     |x|^{4h_k}
     \lng \si_{f_a}^{k}(x,\bar x)\si_{f_b}^{l}(1,1)
     \si_{f_c}^{-(k+l)}(0,0)\rng\ .             \label{yukdef}
\qe
They vanish, unless the three point selection rule is satisfied:
\be
(1-\Th^k)f_a+\Th^k(1-\Theta^l)f_b
-(1-\Th^{k+l})f_c=(1-\Th^k)\tau_a+\Th^k(1-\Th^l)\tau_b\ .
\label{srule3}
\ee
This is to be understood in the sense that the correlator vanishes unless there
are two lattice vectors $\tau_a,\ \tau_b$ such that \req{srule3} holds.

In the calculation of the three twist correlation function we
followed the method described for $k=l$ in \cite{LMN2}. The final result reads

\be
\ds{Y_{f_a,f_b,f_c}^{k,l} =
\lf[ \nu
\prod_{j=1}^{d/2} \Ga_{k_j,l_j}\ri]^{1\ov2}
\sum_{v \in {\cal U}}  e^{-\pi v^t(\Eb_d+2B)Lv}\ ,}
\label{sss}
\ee
where
\be \ba{cll}
{\cal U} &=& \ds{f_{ba} - \tau_{ba} + {1-\Th^{k+l}\ovx 1-\Th^\phi}\La\ ,}\nnn
\ds{\Ga_{k_j,l_j}} &=& \left\{ \ba{c@{\quad:\quad}l}
{\ds{\Ga(1-k_j)\Ga(1-l_j)\Ga(k_j+l_j)}
\ov\ds{\Ga(k_j)\Ga(l_j)\Ga(1-k_j-l_j)}} &
 \ds{0<k_j+l_j<1} \\
 \noalign{\vspace{0.3cm}}
\ds{\Ga_{1-k_j,1-l_j}} & \ds{1<k_j+l_j<2\ ,}\ea\right.
\ea \label{abbrev}
\ee
and $L$ is a block diagonal matrix the $j$-th diagonal
block of which is given by

\be
\ba{clccccr}
    L_j &=&  {\ds{1}\ov\ds{|\lambda_j|}}
   \lf( \Eb_2+ i\,{\rm sgn}(\lambda_j\ri) \epsilon) &,& \epsilon
    = \lf( \ba{clcr} 0 & 1 \nnn -1 & 0 \ea \ri) \, ,  \nnn
    \ds{\la_j} &:=&
    \ds{\cot(\pi k_j)+\cot(\pi l_j)}
    &,& \ds{j=1,\ldots,d/2\ .}
\ea
\ee
%defined by
%\be \ba{clclcr}
%     \ds{F_{2j-1,2j-1}} &=& \ds{F_{2j,2j}} &=& \ds{\fc{1}{\la_j}\ ,} \nnn
%     \ds{F_{2j-1,2j}} &=& \ds{-F_{2j,2j-1}} &=& \ds{\fc{i}{\la_j}\ ,} \ea \ee
%   $$\ds{\la_j :=\cot(\pi k_j)+\cot(\pi l_j)\ , \ \ \ j=1,\ldots,d/2\ .}$$

One can demonstrate, as we did below
\req{range2},
that the summation range ${\cal U}$ does not depend on the particular choice of
$\tau_a$ and $\tau_b$.

For $k=l$ \req{sss} reduces to results known before \cite{DFMS,EJL}. Here
the previous criterion applies again: the results of
\cite{BKM} are correct if \req{det} holds.
If this condition is violated the instanton sum in \req{sss}
becomes clearly restricted compared to the results of \cite{BKM}.
Our formulae, with $B$ set to nought, agree
however with the results of \cite{CGM}: In the cases
inspected there one either has a prime twist order $N$ or
$k$ divides an element of $l+N\ZZ$.

\sect{Duality invariance}

When dealing with
discrete symmetries which act on the
background moduli the lattice basis is somewhat more
convenient to work with.
To have a notation free of indices
we combine its $d$ elements into
the columns of a quadratic matrix $e$. Consequently the metrical and the
antisymmetric background tensor read in the lattice basis $g= e^t G e$,
$b= e^t B e$, respectively. We indicate
 the lattice basis counterpart
of a vector $v$ by a hat ($v=e\hat v$).
Since the twist is  an automorphism of $\Lambda$
we have $\Th e= eQ$ ($Q$ integer). This leads to another
version of the background consistency conditions mentioned in
section 2:
\be  Q^t(g+b) Q = (g+b)\ . \label{modul} \ee

If the $Z_N$ orbifold model is to be duality invariant the
couplings \req{svs}, \req{sss} will provide clues concerning the action
of duality on the twisted sector ground states.
Since for orbifolds in $d>2$ the generators of target space symmetries
are not completely known, we will focus our attention on the class
of duality maps described in \cite{MS1,MS2}.
The ``canonical'' involutive duality operation simply inverts the
background matrix $(g+b)$. For the image $(g+b)^{-1}$ the property
\req{modul} is usually lost.
To remedy this defect one composes the canonical duality transformation
with the elements $W$ of a family of ${\rm GL}(d;\ZZ)$ transformations
for which ($Q^* := (Q^t)^{-1}$)
\be W\,Q\,=\,Q^* W \ee
is required.
This gives rise to the following
family of duality transformations
\cite{MS1}:
\bea
\ds{S(g)} &=&  \ds{W \igpb g \igmb W^t\ ,} \nnn
\ds{S(b)} &=& \ds{- W \igpb b \igmb W^t\ ,}\ \nnn
 \ds{S(e)} &=& e{\ds{1}\ov\ds{g-b}}W^t \ .
\label{wdu}
\eea
These transformations are modular, i.e., $S(g+b)$ satisfies \req{modul}.

One can show that for the
$\lng V\sigma^{-k}\sigma^k\rng$ couplings to be preserved, the twist fields
must transform according to
\eq
{\sigma}^{k}_{\f_{q}} \mapsto \tilde{\sigma}^{k}_{\f_{q}}=
\sum_{r=1}^{N_k} U_{qr}^{(k)} {\sigma}^{k}_{\f_{r}}\ ; \quad
N_k = \det(1-Q^k)\ ,
\label{tdu}
\qe
with
\eq
U_{qr}^{(k)} = \frac{\omega_k(g,b)}{\sqrt{N_{k}}}
\ e^{2\pi i\f_{r}^{t} (1-{Q^{\ast}}^k) W \f_{q}}\ .  \label{uk}
\qe
Eq. \req{ssv} does not determine the
%also allows for an additional
phase factor
$\omega_k(g,b)$.
Note that the matrix $U^{(k)}$ depends on $W$.
Whenever $\f_r$ is not twist invariant
%For $Q$--orbits of $\f_r$ whose length exceeds the trivial value 1
(a case which usually occurs in higher twisted sectors)
the right hand side of \req{uk} must be summed over the
complete $Q$--orbit ${\cal O}_{\f_r}$ of $\hat f_r$.

It may be checked that all correlation functions are invariant under the
duality transformation consisting of \req{wdu} and \req{tdu}. As to the Yukawa
 couplings the calculation
is somewhat more involved \cite{ejss}, and here we would only like to stress
the
decisive role played by the
{\em restricted} summation range in the instanton series
in \req{sss}. The statement of duality invariance can be cast into
\eq
\tilde Y_{\f_a,\f_b,\f_c}^{k,l} [S(g+b)] =
       Y_{\f_a,\f_b,\f_c}^{k,l} [g+b] \ ,
\qe
where the tilde points out that the ``rotated" twist fields \req{tdu}
must be inserted into \req{yukdef}.
It turns out that this condition can be satisfied
upon choosing
\eq \omega_{k}(g,b) =
\prod_{\mu=1}^{M}
\lf[\frac{{\det_\mu} (GZ_\mu-iB)} {{\det_\mu}(GZ_\mu+iB)}\ri]^{-\frac{1}{4}
(1-2k_\mu)}\ ,
\qe
Here the product runs over the various complex target subspaces
${\cal T}_\mu$ $(1\le\mu\le M)$
for which the twist acquires the eigenvalues $u_{\mu}$ or $\bar u_{\mu}$.
A determinant restricted to ${\cal T}_\mu$ is referred to as $\det_\mu$
and $Z_\mu$ is a block diagonal
matrix in  ${\cal T}_\mu$
whose $2\times2$ dimensional blocks are $\epsilon$
$(-\eps)$ if the twist eigenvalue
of the associated complex direction reads $u_\mu$ ($\bar u_\mu$).

It is sometimes useful to have a formula for the Yukawa
coupling evaluated at the dual background. From the above it follows that this
is given by
\eq
Y_{\f_a,\f_b,\f_c}^{k,l} [S(g+b)] = \sum_{p=1}^{N_k}
\sum_{q=1}^{N_{l}}
\sum_{r=1}^{N_{k+l}} U^{(k)\dag}_{ap} U^{(l)\dag}_{bq} U^{(k+l)}_{cr}
Y_{\f_p,\f_q,\f_r}^{k,l} [g+b]\ . \label{char}
\qe
The previous formula can in general be further simplified
owing to the following two facts: (i) $Y^{k,l}_{\f_p,\f_q,\f_r}$
vanishes if the selection rule (\ref{srule3}) ceases to hold.
(ii) \newcommand{\s}{\hat s}
The Yukawa couplings are invariant under certain discrete translations
$X\mapsto X+2\pi f$ of the target space. We must then ask whether
the induced opposite shifts of the fixed point representatives
$\f_p$, $\f_q$, $\f_r$ are well--defined.
It turns out that $f$ must
be a fixed point, say, from the $\varphi$--th sector.
In addition, it follows that
that $\varphi$ has to divide $\phi$.

\sect{Conclusion}

In section 2 we sketched  how to calculate the classical part of the four twist
 field correlator via the path integral method.
No restrictions were made concerning
%the antisymmetric background tensor or
the choice of the two twist sectors labelled by $k$, $l$,
%respectively.
and a general twist compatible background was considered.
Generically it is not sufficient
to identify the classical instanton solutions by reading off
the global monodromy cosets contributed by the standard pair $C_1$, $C_2$
of two independent zero net twist loops. It is important to further
exclude those solutions which are displaced by a non--zero lattice vector
upon being transported along a closed contour with all twist fields
located in the interior.

Whereas this requirement was automatically satisfied for the cases
considered in \cite{DFMS,HV} it is no longer {\em a priori} the case
for the more general setting of \cite{BKM}.
We had to introduce an additional sector index $\phi$ given by
the greatest common divisor of the sectors $k$, $l$ and of the twist
order $N$. As can be read off from (\ref{sss}) and
(\ref{abbrev})
$\phi$ determines which holomorphic {\em or}
antiholomorphic instantons will contribute to
the moduli dependent part of a Yukawa coupling.
Although the duality map reshuffles the moduli in a nontrivial way,
both the Yukawa couplings and the annihilation couplings
turn out to be invariants \cite{ejss}. This property strongly
supports our reasoning.

\vskip 2cm
\centerline{\bf Acknowledgements}
\vskip 0.5cm

We would like to thank Jens Erler, Albrecht Klemm and
Hans--Peter Nilles for many useful
discussions.

\end{document}